\documentclass[11pt,a4paper,english,nofootinbib]{revtex4}
\usepackage{lmodern}

\usepackage[T1]{fontenc}
\usepackage[latin9]{inputenc}
\usepackage{babel}
\usepackage{amsmath}
\usepackage{amssymb}
\usepackage{esint}
\usepackage{graphicx}
\usepackage[unicode=true,pdfusetitle,
 bookmarks=true,bookmarksnumbered=false,bookmarksopen=false,
 breaklinks=false,pdfborder={0 0 1},backref=false,colorlinks=false]
 {hyperref}

\makeatletter

\newcommand{\lyxmathsym}[1]{\ifmmode\begingroup\def\b@ld{bold}
  \text{\ifx\math@version\b@ld\bfseries\fi#1}\endgroup\else#1\fi}

\@ifundefined{textcolor}{}
{%
 \definecolor{BLACK}{gray}{0}
 \definecolor{WHITE}{gray}{1}
 \definecolor{RED}{rgb}{1,0,0}
 \definecolor{GREEN}{rgb}{0,1,0}
 \definecolor{BLUE}{rgb}{0,0,1}
 \definecolor{CYAN}{cmyk}{1,0,0,0}
 \definecolor{MAGENTA}{cmyk}{0,1,0,0}
 \definecolor{YELLOW}{cmyk}{0,0,1,0}
}

\usepackage{latexsym}
\usepackage{bm}

\def\HollowBox #1#2{{\dimen0=#1 \advance\dimen0 by -#2
       \dimen1=#1 \advance\dimen1 by #2
        \vrule height #1 depth #2 width #2
        \vrule height 0pt depth #2 width #1
        \llap{\vrule height #1 depth -\dimen0 width \dimen1} 
       \hskip -#2
       \vrule height #1 depth #2 width #2}}

\makeatother

\begin{document}

\title{Stern-Gerlach Experiment with Higher Spins }

\author{Bayram Tekin }

\email{btekin@metu.edu.tr}

\affiliation{Department of Physics,\\
 Middle East Technical University, 06800, Ankara, Turkey}

\date{\today}
\begin{abstract}
We analyze idealized sequential Stern-Gerlach experiments with higher spin particles.  This analysis serves at least two  purposes: The widely discussed spin-$1/2$ case leads to some misunderstandings since the probabilities are always evenly distributed for the sequential orthogonal magnets which does not generalize to higher spins. A detailed discussion of the  higher spin case, as is done here, is highly useful. Secondly,  the Wigner rotation matrices for generic spins become conceptually more transparent with this physical example. We also give compact formulas for the probabilities in terms of the angle between the sequential SG apparatuses  for generic spins.  We work out the spin-$1/2$, spin-$1$ and spin-$2$ cases explicitly. Since there are some confusing issues regarding the actual experiment, we also compile a "facts and fiction" section on the  Stern-Gerlach experiments. 
\end{abstract} 

\maketitle

\section{Introduction}
We begin with the following concrete problem:
\,\,{\it Given two idealized  sequential SG magnets that point in two fixed  directions  as shown in Figure 1, what are the probabilities of possible outcomes from the second magnet for spin-$s$ particles in terms of $s$ and the angle between the 
magnets  ? \ } 
\begin{figure}[h]
\includegraphics[width=8 cm]{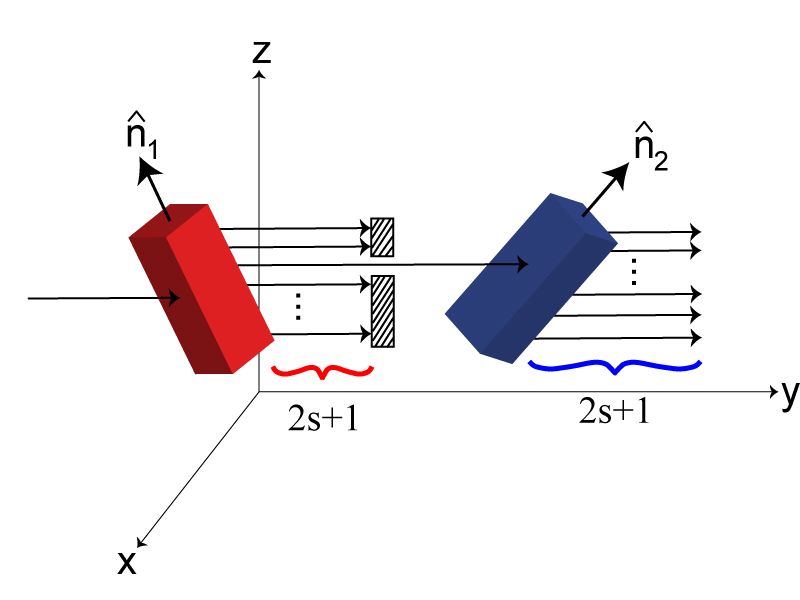}
\caption{Schematic representation of  two sequential Stern-Gerlach (SG) magnets.  An unpolarized  beam of neutral  spin-$s$ particles with magnetic dipole moments, coming from the oven located at the far left,  enter the first magnet  whose magnetic field is pointing in the $\hat{n}_1$ direction, they are split into $2 s+1$ states and only one of these states is allowed to enter the second magnet, the rest are blocked. The second magnet is pointing in the $\hat{n}_2$ direction. On the far right a detector detects the split atoms.
 The question is to write the probabilities of all possible  outcomes in terms of the angle between the magnets, or in short in terms of $\hat{n}_1 \cdot \hat{n}_2$. }
\end{figure}

In the wording of the problem, the reservation {\it idealized} is quite important: As we shall discuss at the end, the actual SG experiment is  beautiful  but a very complicated one and the final probabilities cannot simply be explained {\it  exactly}  in terms of classical quantities.  With this reservation in mind, let us note one of the probabilities here. Suppose we block all but the highest spin (measured in the  $\hat{n}_1$ direction) after the first magnet. Let us call this state to be the $ | \hat{n}_1; s \rangle $ state. Then  this state enters the second magnet and the probability of getting  the $ | \hat{n}_2; s \rangle $ state out of $ | \hat{n}_1; s \rangle $ state is
\begin{equation}
P_{s\hbar \rightarrow s\hbar}  \equiv  | \langle \hat{n}_2; s | \hat{n}_1; s \rangle|^2=  \frac{1}{2^{2 s}}\Big(1+ \hat{n}_1 \cdot \hat{n}_2 \Big)^{2 s}. 
\label{max}
\end{equation}
A priori, one might have  guessed the angular dependence but the dependence on the spin cannot be easily guessed. As we shall see below, in general, when one looks for other probabilities of the form $P_{m_1 \hbar \rightarrow m_2\hbar} $, the problem becomes more involved. Before we start the computations properly, let us say a few words on the history and the importance of the experiment, as well as the people involved and also note what is left somewhat unclear by the canonical case of spin-$1/2$.

\subsection{A Pinch of History}

Stern-Gerlach (SG) experiment \cite{SG} is a very useful tool in teaching many aspects of quantum mechanics from the notion of spin of particles (or atoms) to superposition states, the problems of weak and strong measurements, decoherence, entanglement and density matrix formalism etc.  Bernstein goes  so far as to say "{\it ... all of quantum mechanics is summarized in the Stern-Gerlach experiments-at least all of quantum mechanics that is really mysterious}"\cite{Bernstein}. The idea of the experiment is simple: In an inhomogeneous magnetic field, with a sufficient gradient, a mixed collection of neutral atoms are expected to split  into various groups due to their possible  magnetic dipole moment, which is quantized in quantum mechanics. The actual execution of the experiment and understanding the outcomes in full detail with quantum mechanics is  not a straightforward task. After one year of experimentation,  Stern and Gerlach published their results in 1922.  Even though what they really measured in the experiment was the magnetic dipole moment of the electron up to 10\%  accuracy, since the notion of electron spin was not known in those days, the correct interpretation of the experiment came only 5 years later \cite{1927} in a highly interesting paper that established the isotropy of some atoms, namely the existence of $s$-orbital electrons. These $s$-orbitals were thought to be non-existent, or unstable,  in the Bohr-Sommerfeld theory. 
As opposed to the actual experiment, the {\it idealized} versions of the experiment  serve as a fascinating teaching tool with many sequential SG apparatuses utilized  to expound upon the idea of superposition states etc.  Many modern books discuss the experiment in varying  details; some even start  teaching quantum mechanics with the experiment.  As an example for undergraduate students, McIntyre's book  \cite{McIntyre} and for graduate students Sakurai \& Napolitano \cite{Sakurai} can be mentioned. 

Many of the books that include the experiment, only discuss the  case of the silver atom, being a spin-$1/2$ system, it is the simplest, perhaps the most elegant case  to discuss that  was used in the original  SG experiment that started all. [Feynman's Lectures on Physics is an exception: Feynman devotes a chapter for the spin-$1$ case \cite{feynman3}.]  Spin-$1/2$ case elegant in the sense that, it shows the quantization of spin (or more properly the magnetic dipole moment)  in the best way possible. But this simple case leaves  many issues unclear if one is interested in issues beyond spin quantization such as superposition states etc. as mentioned above. This is because for the sequential SG experiments where two or more magnets are used to describe various overlaps of states  (as shown in Figure 1), probabilities  are evenly split for the spin-$1/2$ case when the magnets are perpendicular to each other, while this is not the case for higher spins. Let us explain what we mean with an example: Consider the spin up ( $+\frac{ {\hbar}}{2} $ ) silver atoms, described  by the state   $| S_z;+ \rangle $,  coming out of the SG magnets placed along the $\hat{z}$-axis as shown in Figure 2. Take these states and block the spin-down ones. Then, let these spin-up states pass through a second set of magnets in the $\hat{x}$-direction, then 
one expects  to find  half of the atoms to be in the state  $|S_x; + \rangle $ and half in the state  $|S_x; - \rangle $.  (Of course due to symmetry between the $\hat{x}$ and $\hat{y}$ axes, the probabilities  would be again evenly split if one had the second set of magnets lie along the $\hat{y}$-axis.) Most students do not have much trouble in understanding this. 
\begin{figure}[h]
\includegraphics[width=10 cm]{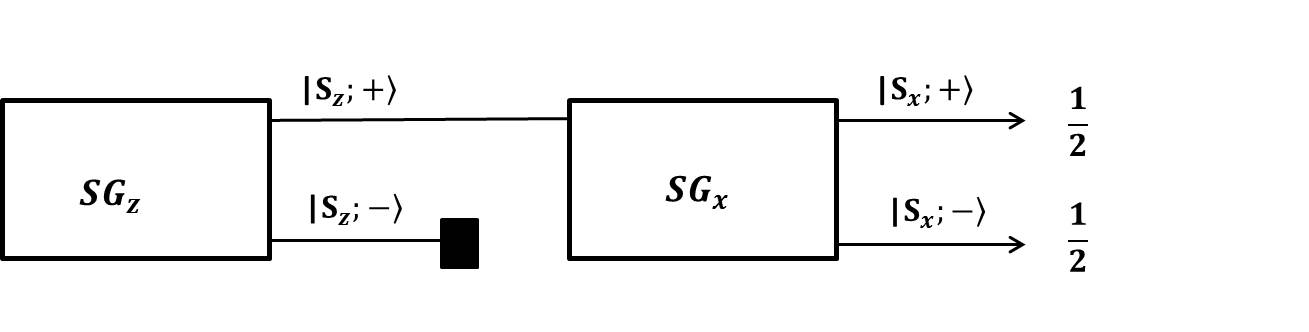}
\caption{A highly simplified picture of two SG magnets that deflect  spin-$1/2$  particles with magnetic dipole moments. The probabilities are evenly distributed for this setting. This is the canonically discussed case to reveal the notion of superposition states. }
\end{figure}

But for the (massive) spin-$1$ case, when they see the following probabilities (some are shown in Figure 3), most of the students feel uneasy:
\begin{equation}
| \langle S_x; \pm 1 | S_z ;1  \rangle |^2 =  \frac{1}{4}, \hskip 0.5 cm | \langle S_x;0 | S_z ;1  \rangle |^2 = \frac{1}{2},   \hskip 0.5 cm | \langle S_x;0 | S_z ;0  \rangle |^2 = 0.
\end{equation} 

\begin{figure}[h]
\includegraphics[width=10 cm]{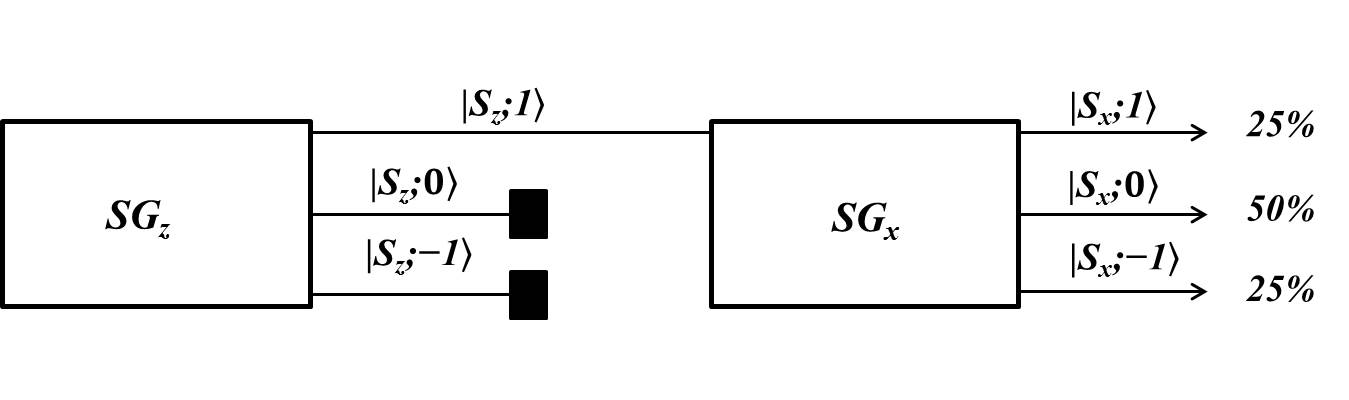}
\caption{The probability out-comes are not evenly distributed  for this orthogonal setting of the magnets, unlike the case of spin-$1/2$. The problem becomes a bit more complicated when the magnets are not orthogonal to each other. Details of the computations of the probabilities are given in the text below for the general orientation case.}
\end{figure}
In the next section we will see how these probabilities are computed, but before that let me note a couple of things about the SG experiment as a teaching tool. Both in my undergraduate and graduate quantum mechanics courses, I spend a fair amount of time  discussing various facets of the experiment for the original silver atoms as well as gedanken higher spin systems. 
I found that the students feel much more comfortable with the discussion of rotations, Wigner matrices etc. when these issues are discussed in the SG context, or at least higher spin SG experiment is given as an example where, for example Wigner matrices appear.  I also found that the history of the original SG experiment with serendipitous events and the fascinating lives of the people involved is so rich that it is one of the best "baits" to attract the students to think about quantum mechanics. Before the lectures on the SG experiment, I assign students to read the dramatic account of the experiment  from the article \cite{cigar}. History of the experiment is not our intention to discuss here, but let us say a few words. Stern and Gerlach's low salaries played an interesting role: they could afford only cheap cigars with high sulfur content which helped  the split-silver atoms  appear in the deposited plates. Born even gave a paid lecture on relativity to help the experiment and solicited funds from a businessman in the States. Long after the experiment, the lives of Stern and Gerlach diverge in such a dramatic way: Gerlach worked for the Germany's war effort, in particular, he became the director of the nuclear energy 
programme (possibly intending  to develop an atomic bomb) and got detained with nine other German scientists, the likes of W. Heisenberg, O. Hahn, M. von Lue, C. F. Von Weizsacker in a house in Cambridge for 6 months after they got captured in operation epsilon  whose scientific advisor was none other than S. Goudsmit, who was one of the originators of the idea of spin.  Stern on the other hand was awarded  the Nobel prize in physics in 1943 with a citation which did not mention the SG experiment but cited Stern's  measurement of the magnetic dipole moment of the proton. Let us now turn to the bulk of the paper where we study idealized SG experiments with higher spins.

\section{ Idealized SG experiments with all spins}

With two sequential SG magnets, the most general probability one needs to compute is the probability of  finding the state  $|\hat{n}_2; m_2 \rangle $  in the state  $|\hat{n}_1; m_1 \rangle$, where of course  $m_{1,2}$ being the components of the spin in the $\hat{n}_{1,2}$ directions, take values from the set $ \{ -s, -s +1 , ..., s-1, s \}$.  So we need to find the result of 
\begin{equation}
P_{m_1\hbar  \rightarrow m_2\hbar } = | \langle \hat{n}_2; m_2 |  \hat{n}_1; m_1 \rangle |^2, 
\end{equation}
in terms of the vectors $\hat{n}_i$ and $s$ as well as  $m_{1,2}$. Before we present the general spin case, let us adopt a very naive approach and carry out the computation for several lower spin cases by brute force, namely by finding the states  $|\hat{n}; m \rangle $, say in vector representation,  and directly computing the probability.

\subsection{Spin-$1/2$  case }
To compute the probability outcomes from the theory what  do we need to know? The answer is: We need to know the angular momentum, in this case, the  spin algebra. So, for the rest of the work this is assumed to be the case. Since this is discussed in almost all textbooks, we shall not dwell on this much. It is also  sort of self-evident: Classical  3 dimensional spatial rotations as applied to a quantum system lead to the commutation relations
\begin{equation}
[ \hat{S}_i, \hat{S}_j] = i \hbar \,\epsilon_{ ij k} \hat{S}_k,
\end{equation}
which are satisfied by  $2 s+1$-dimensional matrices  with  $ s=0,1/2,1,3/2,...$ and the corresponding states are labeled by  the eigenvalues of the quadratic Casimir operator $\hat{S}^2$ together with the eigenvalue of 
our favorite spin component $\hat{S}_z$.  Then without further ado, let us note that the eigenket  of  $\hat{S}_{n}= \hat{S}\cdot \hat{n}$ with eigenvalue $ \hbar/2$ is easily found as
\begin{equation}
|\hat{n}; + \rangle =  \cos  \frac{\theta}{2}  |\hat{z}; + \rangle   + \sin  \frac{\theta}{2}  e^{i \phi },|\hat{z}; - \rangle 
\end{equation} 
where  $\theta$ is the angle measured from the $\hat{z}$ axis in the clockwise direction and $\phi$ is the angle measured from the $\hat{x}$ axis  in the counterclockwise direction. Then, given 
the  vectors 
\begin{eqnarray} 
\hat{n}_i =     \sin\theta_i \cos \phi_i  \,\hat{x} +  \sin\theta_i \sin\phi_i \,\hat{y} +  \cos \theta_i \,\hat{z},
\end{eqnarray}
it is easy to show that one has the following probability 
\begin{equation}
P_{\frac{\hbar}{2} \rightarrow \frac{\hbar}{2} }  = | \langle \hat{n}_2;  + |  \hat{n}_1; + \rangle |^2= \frac{1}{2} ( 1+ \hat{n}_1 \cdot \hat{n}_2 ) 
\end{equation}
Needless to say, one can choose one of the directions , say $\hat{n}_1$, to be in the $\hat{z}$ direction. 
Without a surprise, the "other" probability also follows as (See Figure 4)
\begin{equation}
P_{\frac{\hbar}{2} \rightarrow -\frac{\hbar}{2} }  = | \langle \hat{n}_2;  - |  \hat{n}_1; + \rangle |^2= \frac{1}{2} ( 1- \hat{n}_1 \cdot \hat{n}_2 ), 
\end{equation}
happily yielding $ P_{\frac{\hbar}{2} \rightarrow \frac{\hbar}{2} } +P_{\frac{\hbar}{2} \rightarrow -\frac{\hbar}{2} } =1$.   Let us note that we can define the angle between the two vectors as
\begin{equation}
\cos \varphi \equiv  \hat{n}_1 \cdot \hat{n}_2  =   \sin\theta_1  \sin\theta_2  \cos (\phi_1 - \phi_2 ) + \cos \theta_1 \cos \theta_2.
\end{equation}
\begin{figure}[h]
\includegraphics[width=10 cm]{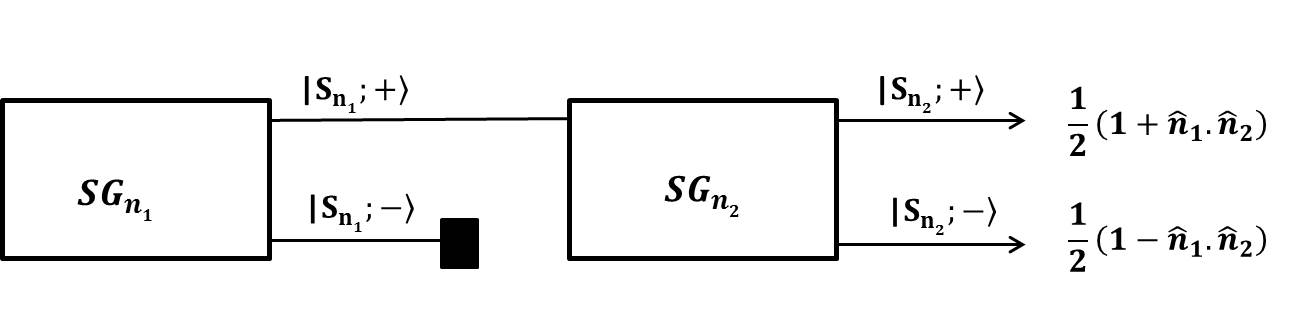}
\caption{Spin-$1/2$ particles enter to two sequential SG magnets that are oriented in generic directions. Probabilities are written on the right. This example serves as a good starting point, but it misses a lot of fine details.}
\end{figure}

This simple discussion was to set the stage, let us now consider the massive spin-1 case (a massless one would have only 2 helicity modes, which we do not discuss here)

\subsection{Spin-$1$  case }
For this case $\hat{S}_n $, is a $3 \times 3$ Hermitian operator with the following eigenkets (up to a phase, as usual)
\begin{eqnarray}
&&|\hat{n}; 1 \rangle =  \cos^2 ( \frac{\theta}{2} ) e^{ - i \phi} |\hat{z}; 1 \rangle   + \frac{1}{\sqrt{2}} \sin \theta \,|\hat{z}; 0 \rangle  + \sin^2 ( \frac{\theta}{2} ) e^{  i \phi}|\hat{z}; -1 \rangle,  \nonumber \\
&&|\hat{n}; 0 \rangle =  -\frac{1}{\sqrt{2}}\sin \theta e^{ - i \phi} |\hat{z}; 1 \rangle   + \cos \theta \,|\hat{z}; 0 \rangle  +\frac{1}{\sqrt{2}} \sin \theta \, e^{  i \phi}|\hat{z}; -1 \rangle,  \nonumber \\
&&|\hat{n}; -1 \rangle =  \sin^2 ( \frac{\theta}{2} ) e^{ - i \phi} |\hat{z}; 1 \rangle   - \frac{1}{\sqrt{2}} \sin \theta \,|\hat{z}; 0 \rangle  + \cos^2 ( \frac{\theta}{2} ) e^{  i \phi}|\hat{z}; -1 \rangle.
\end{eqnarray}
The probabilities can be computed as 
\begin{eqnarray}
&&P_{\hbar \rightarrow \hbar} \equiv |\langle \hat{n}_2; 1 | \hat{n}_1; 1 \rangle|^2 =  \frac{1}{4}(1 + \hat{n}_1 \cdot \hat{n}_2 )^2 , \hskip 0.5 cm P_{\hbar \rightarrow -\hbar} \equiv |\langle \hat{n}_2; -1 | \hat{n}_1; 1 \rangle|^2 =  \frac{1}{4}(1 - \hat{n}_1 \cdot \hat{n}_2 )^2, \nonumber \\
&&P_{\hbar \rightarrow 0} \equiv |\langle \hat{n}_2; 0 | \hat{n}_1; 1 \rangle|^2 =  \frac{1}{2}(1 -(\hat{n}_1 \cdot \hat{n}_2 )^2 ) , \hskip 0.5 cm P_{0 \rightarrow 0} \equiv |\langle \hat{n}_2; 0 | \hat{n}_1; 0 \rangle|^2 = (\hat{n}_1 \cdot \hat{n}_2 )^2.
\label{spin1}
\end{eqnarray}
The others simply follow from these due to symmetry. Observe that probabilities for each given initial state ending to a sum of final states add up to one,  $\sum_{f} P_{i \rightarrow f} =1 $, as needed.  Probabilities noted in Figure 3  follow  from (\ref{spin1}) as a simple example. Leaving the spin-$3/2$ case to the reader as a useful exercise, let us move on to the spin-$2$ case. 

\subsection{Spin-$2$  case }
In this case let us give some more details of the computation, just to show that for arbitrary spin-$s$, this brute force method is simply not practical. 
For spin-$2$, the relevant component of the spin operator in generic $\hat{n}$ direction can be computed from $\hat{S}_n = \hat{S} \cdot \hat{n}$ to get
\[
\hat{S}_n=\hbar \left(
\begin{array}{ccccc}
 2 \cos \theta  & e^{-i \phi } \sin \theta & 0 & 0 & 0 \\
 e^{i \phi } \sin \theta & \cos \theta & \sqrt{\frac{3}{2}} e^{-i \phi } \sin \theta & 0 & 0 \\
 0 & \sqrt{\frac{3}{2}} e^{i \phi } \sin \theta & 0 & \sqrt{\frac{3}{2}} e^{-i \phi } \sin \theta & 0 \\
 0 & 0 & \sqrt{\frac{3}{2}} e^{i \phi } \sin \theta & -\cos \theta & e^{-i \phi } \sin \theta \\
 0 & 0 & 0 & e^{i \phi } \sin \theta & -2 \cos \theta \\
\end{array}
\right),
\]
with, of course, eigenvalues $\{ 2,1,0,-1,-2 \}\hbar$. Let us not write all the eigenvectors, since they are somewhat cumbersome, but give only two of them as examples
\begin{eqnarray}
&&|\hat{n}; 2 \rangle = \Bigg(e^{-2 i \phi } \cos ^4\left(\frac{\theta }{2}\right),e^{- i \phi } \sin \theta \cos ^2\left(\frac{\theta }{2}\right),\sqrt{\frac{3}{8}} \sin ^2 \theta,e^{i \phi } \sin ^2\left(\frac{\theta }{2}\right) \sin\theta ,e^{2 i \phi }\sin ^4\left(\frac{\theta }{2}\right)\Bigg) \nonumber \\
&&|\hat{n}; 0 \rangle = \Bigg(\sqrt{\frac{3}{8}} e^{-2 i \phi } \sin ^2 \theta,-\sqrt{\frac{3}{8}} e^{-i \phi } \sin 2 \theta,\frac{1}{4}(3 \cos 2 \theta+1),\sqrt{\frac{3}{8}} e^{i \phi } \sin 2 \theta , e^{2 i \phi }\sqrt{\frac{3}{8}} \sin ^2 \theta\Bigg). \nonumber
\end{eqnarray}
As a specific case, as shown in Figure 5,  let us block all the states except $|\hat{n}_1; 0 \rangle$ after the first magnet and compute the probability of obtaining various states for this particular  experiment  for the second magnet.
\begin{figure}[h]
\includegraphics[width=10 cm]{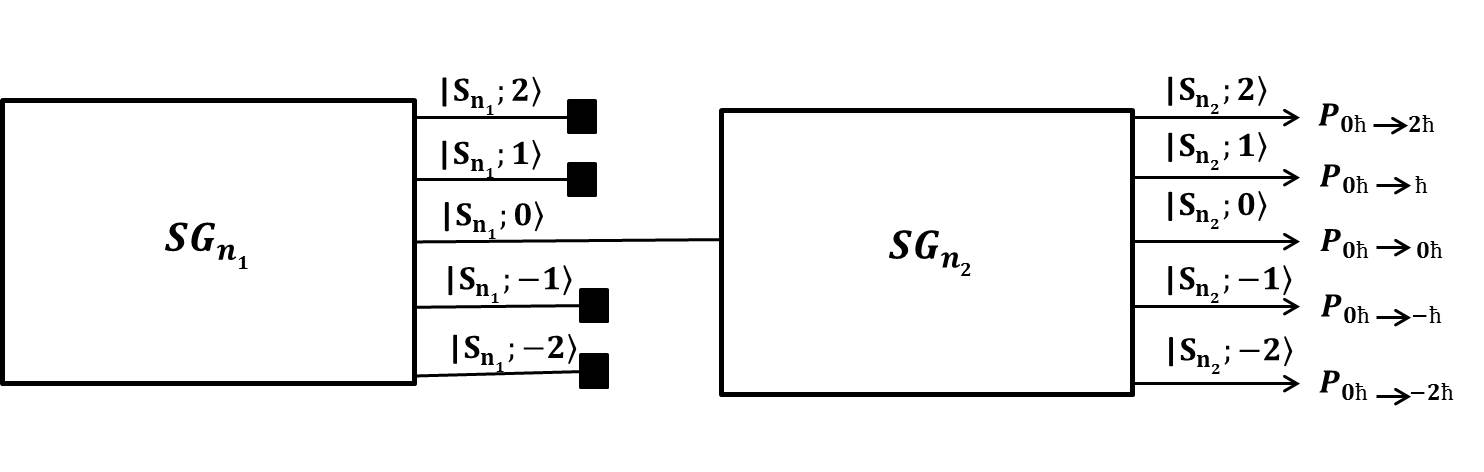}
\caption{Two sequential SG magnets in arbitrary orientations. Probabilities are calculated in the text. This is a nice example which is still manageable by hand using the  brute force techniques, but for any higher spin, one should consult the formalism given under the subsection "Spin-$s$ case"  }
\end{figure}

Using the explicit forms of the state vectors one arrives at  
\begin{eqnarray}
&&P_{0 \hbar \rightarrow  \pm 2\hbar} \equiv  |\langle \hat{n}_2;  \pm 2 |\hat{n}_1; 0 \rangle|^2 =   \frac{3}{8}\Big (1- (\hat{n}_1 \cdot \hat{n}_2 )^2 \Big)^{2} \nonumber \\
&& P_{0\hbar \rightarrow 0\hbar} \equiv  |\langle \hat{n}_2; 0 |\hat{n}_1; 0 \rangle|^2  =   \frac{1}{4}\Big (1- 3(\hat{n}_1 \cdot \hat{n}_2 )^2 \Big)^{2}.
\end{eqnarray}
One can also easily show , due to symmetry and the fact that the total probability is unity that 
\begin{equation}
P_{0\hbar \rightarrow \pm\hbar}  \equiv  |\langle \hat{n}_2;  \pm 1 |\hat{n}_1; 0 \rangle|^2  =  \frac{3}{2}(\hat{n}_1 \cdot \hat{n}_2)^2\Big (1-  (\hat{n}_1 \cdot \hat{n}_2 )^2 \Big).
\end{equation}

 We now turn our attention to the general spin case and answer the question posed in the beginning of the paper. 
\subsection{Spin-$s$  case } 
Our task now is to compute the following probability 
\begin{equation}
P_{m_1\hbar  \rightarrow m_2\hbar } = | \langle \hat{n}_2; m_2 |  \hat{n}_1; m_1 \rangle |^2, 
\end{equation}
whose meaning is clear when referred to  Figure 1. Not to clutter the notation, and since the final result is independent of the choice of coordinates anyway, let us choose $ \hat{n}_1 = \hat{z}$, which then reduces our task to the following computation
\begin{equation}
P_{m_1\hbar  \rightarrow m_2\hbar } = | \langle \hat{n}_2; m_2 |  \hat{z}; m_1 \rangle |^2.
\end{equation}
This is all rather obvious, the crucial step is the next one: We can obtain  $| \hat{n}_2; m_2 \rangle$ from   $| \hat{z}; m_2 \rangle$ by two rotations: First rotate about the $\hat{y}$ axis by an angle $\theta$, then rotate  about the $\hat{z}$ axis by an angle $\phi$  (both in the counter-clockwise direction) as 
\begin{equation}
 | \hat{n}; m_2 \rangle  = e^{ -\frac{i}{\hbar} \hat{J}_z \phi }\, e^{ -\frac{i}{\hbar} \hat{J}_y \theta}    | \hat{z}; m_2 \rangle,
\end{equation}
which further reduces our computation to 
\begin{equation}
P_{m_1\hbar  \rightarrow m_2\hbar } =  | \langle \hat{z}; m_2 |  e^{ \frac{i}{\hbar} \hat{J}_y \theta}\,   e^{ \frac{i}{\hbar} \hat{J}_z \phi }          | \hat{z}; m_1 \rangle |^2  \equiv | d^{(s)}_{m_2, m_1} (-\theta) |^2,
\end{equation}
where the angle $\phi$ drops out and we have defined 
\begin{eqnarray}
d^{(s)}_{m_2, m_1} (\theta)  \equiv \langle \hat{z}; m_2 | e^{ -\frac{i}{\hbar} \hat{J}_y \theta}  |  \hat{z}; m_1 \rangle,
\end{eqnarray}
which are the elements of little-$d$ (Wigner)  rotation matrices.  Since, the computations of these objects are well covered in many books (for example, see \cite{Sakurai}), I just quote the final result here
\begin{eqnarray}
d^{(s)}_{m_2, m_1} (\theta) =\sum_k (-1)^{ k-m_1 + m_2} \Gamma^s_ { m k} \times \Big (\cos \frac{ \theta}{2} \Big)^{ 2 s - 2k +m_1 - m_2}  \Big (\sin \frac{ \theta}{2} \Big)^{ 2k -m_1 + m_2},
\end{eqnarray}
where  the capital gamma stands for 
\begin{equation}
 \Gamma^s_ { m k}  \equiv  \frac{\sqrt{ (s+m_1)! (s-m_1)! (s+m_2)! (s-m_2)!}}{ k! (s+m_1-k)!(s-k-m_2)! (k-m_1+m_2)!},
\end{equation}
and the sum runs over all possible integer $k$ values that do not yield negative arguments of the factorials. Since in this choice of coordinates, $\theta$ is the angle between the two magnets, we can write the final result in a coordinate-independent way as
\begin{eqnarray}
d^{(s)}_{m_2, m_1} (\theta) = \frac{1}{2^s} (1+ \hat{n}_1 \cdot \hat{n}_2 )^{s} \Big (\frac{1+ \hat{n}_1 \cdot \hat{n}_2 }{1- \hat{n}_1 \cdot \hat{n}_2 } \Big )^{\frac{m_1-m_2}{2}}
\sum_k (-1)^{ k-m_1 + m_2} \Gamma^s_ { m k}  \Big (\frac{ 1- \hat{n}_1 \cdot \hat{n}_2 }{1+ \hat{n}_1 \cdot \hat{n}_2 } \Big)^{k}.
\end{eqnarray}
This is the amplitude of measuring  the spin component to be $m_2 \hbar$ in the second magnet oriented along $\hat{n}_2$ if we allow only $m_1 \hbar$ spins from the first magnet oriented along $\hat{n}_2$ as depicted in Figure 1.  From this amplitude one computes the desired probability, such as the expression (\ref{max}). When $m_1 = m_2 = s$,  one has  $\Gamma^s_ { m 0} =1 $ and one obtains  the amplitude as 
\begin{equation}
d^{(s)}_{s, s} (\theta) = \frac{1}{2^s} \Big (1+ \hat{n}_1 \cdot \hat{n}_2 \Big )^{s}. 
\end{equation}
 Furthermore, as this amplitude is the most general one needed, one can put as many sequential SG magnets as one desires to compute the various probability out comes.  Let us now collect 
some facts and fiction regarding the SG experiments and conclude.

\section{Facts and Fiction in the SG experiments } 
\begin{itemize}
\item {\it  Stern and Gerlach were trying to prove the quantization of the spin of the electron.} No, they knew nothing about the spin of the electron. They were trying to (dis)-prove "space quantization", that is the quantization of the orbital angular momentum as suggested by the Bohr-Sommerfeld theory, which was then the theory of the atom in old quantum mechanics.  In fact, when they observed the splitting, they celebrated Bohr with, now a famous postcard, that shows the photo of the splitting.  Of course, from today's vantage point, they had the wrong theory in their hands, because the lowest non-trivial angular momentum is $\ell =1$ which would lead to 3 lines instead of the 2 they observed. Why then, did they celebrate Bohr? Because, Bohr as well as others did not think that zero angular momentum was stable, so, they eliminated the $m_\ell =0$  component \cite{Bernstein}.

\item {\it  After the SG experiment, surely, people must have realized that there should be  another quantum number in the atom besides the 3 already known at the time?} This is also not correct, as partially answered above. The experiment did not actually play much role  in the introduction of the notion of the spin quantum number. What played role is the spectrum  of the atoms in a magnetic field. More specifically the duplexity of the lines due to the (anomalous) Zeeman effect. Pauli suggested that there must be a fourth quantum number without giving a related motion to the electron, Kronig (as a graduate student ) noted that the duplexity could be due to the spinning motion of the electron, but he was scared off the idea by  the "elders", who earlier entertained the idea of a rotating electron about an axis passing through its center only to find that such a motion is inconsistent with special relativity as some points on the electron must rotate with speeds two orders of magnitude faster than the speed of light.  Uhlenbeck and Goudsmit (also graduate students)  suggested the same notion of spinning electron to explain the duplexity of the spectrum. They were also scared off, but their paper was already published by the time they wanted to retract. What was Pauli and Bohr thinking about the fourth quantum number?.  What kind of physical motion or phenomenon did they attach to it ? This is actually an important point: They thought that the electron does not have an independent, intrinsic property,  call it spin or some other number, the  fourth quantum number arises, just like the other 3 (quantized energy, quantized angular momentum and quantized component of angular momentum),  according to them, when the electron is  bound to the atom.  Who could blame them? Nobody has attached a physical intrinsic quantity to an electron besides its mass and charge at the time.  Dirac's discovery of the relativistic equation for the electron and his prediction of the duplexity and the gyromagnetic factor supported Bohr and  Pauli, because,  if the notion of spin were intrinsic why should it appear in the relativistic theory which has something to the motion \cite{flying}?  Hence comes the topic of next item. 

\item {\it There is no way to measure the spin of an electron in a SG type experiment} No this is not correct, even though this has become somewhat of a folklore because of the obvious dominance of the Lorentz force over the force on the magnetic dipole moment of the electron.
 But it is a rather interesting issue, with a remarkable history \cite{flying}: Bohr always maintained that due to uncertainty  there is no way to measure the spin of an isolated electron and Pauli agreed and in fact wrote a review trying to argue against the published suggestions for experiments, such as the longitudinal SG experiments  where electrons are basically expected to split in a race according their dipole moment in a magnetic field which is inhomogeneous along the direction of their motion.  In principle this is doable as was suggested in \cite{electrons} with the methods invented by the Nobel prize winner  H. Dehmelt.  See also the related notion of "continuous SG experiments"  \cite{cont} that lead to a very accurate determination of the gyromagnetic factor of the electron. 

\item {\it The strength of the gradient of the magnetic field is not important.} This is not correct at all.  Unfortunately, this wrong  view comes from utilizing the  idealized version of the experiment as a teaching tool, as we did above. An honest computation of the predictions  of probabilities must make use of the multi-component Schr\"{o}dinger equation and its wave-packet solutions. 
Consider just the case of the silver atom (that is the spin-$1/2$ case). The atoms coming out of the oven,which is around 1000 Celsius, have a Maxwell-Boltzmann distribution with velocities centered around 1km/s.  They form an unpolarized ensemble  even after they are collimated. We must then, consider a mixed ensemble of particles, interacting with a SG magnet for about $10^{-4}$ seconds (as was the case in the original SG experimental set up)   and getting detected at the screen.  Of course, this is quantum mechanics and one does not really have trajectories. What happens is that the initial superposition states that form the wave packet evolve in time in such a way that  spin-up and spin down parts of the amplitude are peaked around different points in the $\hat{z}$ direction, due to the magnetic field-dipole interactions. Of course, depending on the gradient of the magnetic field, this splitting may not lead to a measurement. In fact, the notion of weak measurement is essentially realized by this idea of having magnetic dipoles interact with magnetic fields with weak gradients \cite{ahar}. I suggest the reader to work out how the centers of the wave-packet will change in the lowest order approximation (namely,  assuming that the magnetic field can be expanded in Taylor series and the resulting spinor Schr\"{o}dinger equation splits into two equations, where the problem reduces to the analogous problem of motion under constant "gravity" ).  The reader will also notice that the lip-like structure in the actual experiment can be explained with the help of two things: for the consistency of the the no monopole condition $\vec{\nabla} \cdot \vec{B}=0$ , the magnetic field must change in another direction, hence for some atoms, the measurement is actually not in the intended direction but in the perpendicular direction. Also, as noted there are no clean-cut trajectories, smearing out is expected. 
 
\item{ \it Silver atom was a random choice in the experiment}  Yes and no. Stern was the first person to measure directly the speed of gas molecules in 1920 using the silver vapor \cite{beams}. (It is amazing that it took such a long time to realize this mid 18th century idea.)  Of course heavier atoms go slower and so their speeds are easier to measure. So, Stern was already familiar with the silver atom. The fact that silver has 46 electrons forming a closed shell  with zero angular momentum and the 47th electron in the $5s$ state with zero orbital angular momentum worked well for Stern and Gerlach. But, in retrospect, with their beautiful experiment  they lent support for the wrong theory \cite{wrong}

\end{itemize}

\section{Conclusions}

We have studied the Stern-Gerlach experiment for generic spin-$s$ particles in the idealized setting to hopefully remove some of the possible  misunderstandings introduced in the canonically-discussed spin  one-half case. We have also given compact formulas for the probability outcomes  which boil-down to the elements of the Wigner rotation  matrices. As explicit examples,  we have worked out the details of spin-$1/2$, spin-$1$ and spin-$2$ for generic orientations of two sequential SG magnets. Of course this discussion could be extended to any number of sequential magnets. We have compiled a facts and fiction section which we hope would serve to remove some unclear points in the idealized version of the SG experiments. 

\section{Acknowledgments}
This work is partially supported by  T\"{U}B\.{I}TAK  grant 113F155. I would like to thank  Nur \"{U}nal and Deniz Tekin for their kind help with the figures.

\end{document}